\documentclass[a4paper,fleqn,usenatbib]{mnras}

\usepackage{mathptmx}
\usepackage[T1]{fontenc}
\usepackage{ae,aecompl}

\usepackage{graphicx}	% Including figure files
\usepackage{amsmath}	% Advanced maths commands
\usepackage{amssymb}	% Extra maths symbols
\usepackage{gensymb}	% Extra maths symbols 
\usepackage{natbib} %Biblography 
\title[Analysis of a solar pore]{Analysis of a spatially deconvolved solar pore}

\author[C. Quintero Noda et al.]{C. Quintero Noda,$^{1}$\thanks{E-mail: carlos@solar.isas.jaxa.jp}
T. Shimizu,$^{1}$
B. Ruiz Cobo,$^{2,3}$
Y. Suematsu,$^{4}$
\newauthor
Y. Katsukawa,$^{4}$
K. Ichimoto,$^{5}$
\\
$^{1}$Institute of Space and Astronautical Science, Japan Aerospace Exploration Agency, Sagamihara, Kanagawa 252-5210, Japan\\
$^{2}$Instituto de Astrof\'isica de Canarias, E-38200, La Laguna, Tenerife, Spain.\\
$^{3}$Departamento de Astrof\'isica, Univ. de La Laguna, La Laguna, Tenerife, E-38205, Spain\\
$^{4}$National Astronomical Observatory of Japan, 2-21-1 Osawa, Mitaka, Tokyo 181-8588, Japan\\
$^{5}$Kwasan and Hida Observatories, Kyoto University, Kurabashira Kamitakara-cho, Takayama-city, 506-1314 Gifu, Japan\\
}

\date{Accepted XXX. Received YYY; in original form ZZZ}

\pubyear{2016}

\begin{document}
\label{firstpage}
\pagerange{\pageref{firstpage}--\pageref{lastpage}}
\maketitle

\begin{abstract}
Solar pores are active regions with large magnetic field strengths and apparent simple magnetic configurations. Their properties resemble the ones found for the sunspot umbra although pores do not show penumbra. Therefore, solar pores present themselves as an intriguing phenomenon that is not completely understood. We examine in this work a solar pore observed with Hinode/SP using two state of the art techniques. The first one is the spatial deconvolution of the spectropolarimetric data that allows removing the stray light contamination induced by the spatial point spread function of the telescope. The second one is the inversion of the Stokes profiles assuming local thermodynamic equilibrium that let us to infer the atmospheric physical parameters. After applying these techniques, we found that the spatial deconvolution method does not introduce artefacts, even at the edges of the magnetic structure, where large horizontal gradients are detected on the atmospheric parameters. Moreover, we also describe the physical properties of the magnetic structure at different heights finding that, in the inner part of the solar pore, the temperature is lower than outside, the magnetic field strength is larger than 2~kG and unipolar, and the LOS velocity is almost null. At neighbouring pixels, we found low magnetic field strengths of same polarity and strong downward motions that only occur at the low photosphere, below the continuum optical depth $\log \tau=-1$. Finally, we studied the spatial relation between different atmospheric parameters at different heights corroborating the physical properties described before. 
\end{abstract}

\begin{keywords}
Sun: photosphere -- Sun: surface magnetism -- techniques: polarimetric
\end{keywords}

%%%%%%%%%%%%%%%%%%%%%%%%%%%%%%%%%%%%%%%%%%%%%%%%%%

%%%%%%%%%%%%%%%%% BODY OF PAPER %%%%%%%%%%%%%%%%%%

\section{Introduction}

The solar photosphere is characterized by a wide variety of magnetic structures of different spatial size, complexity and magnetic field strengths, ranging from sunspots, pores, and network to small internetwork bright points. The morphology of sunspot umbrae and pores is very similar what produces that the second one is sometimes called ``naked umbrae'' \citep{Solanki2003}. The reason behind is that pores, as the inner part of sunspots, are detected in white light observations as dark regions where the photospheric convection and plasma motions have been inhibited due to the presence of a strong magnetic field \citep[see, for instance,][]{Biermann1941,Cowling1953}. 

Solar pores present themselves with diameters of a few Mm, much smaller than the usual size of a sunspot (up to 40 Mm), a field strength in the order of 2 kG, and a lifetime of typically less than a day. Moreover, the absence of filamentary penumbra surrounding pores suggests a simpler magnetic configuration, that can be  represented by a magnetostatic flux tube with a predominantly vertical magnetic field \citep{Simon1970}. However, high spatial-resolution observations indicate that some pores contain a wide variety of fine bright features, such as bright dots or light bridges that may be signs of a convective energy transportation mechanism \citep[][]{Sobotka1999,Keil1999,Hirzberger2003,Giordano2008,Sobotka2013}. For additional information about the general properties of solar pores and sunspots, see the reviews of \cite{Sobotka2003}, \cite{Solanki2003}, \cite{Thomas2004} or a more modern one as \cite{Borrero2011}.

Recent studies like the one performed by \cite{Sobotka2012} examined the magnetic properties of pores in detail, performing inversions of the full Stokes vector in order to infer the atmospheric physical parameters. They observed downflows at the borders of the structure, in agreement with \cite{Keil1999}, \cite{Hirzberger2003}, \cite{Morinaga2008}, and \cite{Scharmer2011}, traces of the magnetic field with the shape of spines extending more than $3.5$ arcsec from the white light pore borders, similar to the results of \cite{Cameron2007}, and a linear correlation between the temperature and the vertical component of the magnetic field, with lower temperatures in the regions of stronger magnetic fields. 

Consequently, the work of \cite{Sobotka2012} provides a big step in the understanding of the nature of solar pores although there are some limitations in their analysis. We aim to surpass these limitations in the present work aiming to complement their findings. One of the limitations is the configuration used in the inversion of the Stokes profiles, with constant stratifications with height for the atmospheric parameters, except the temperature that was perturbed as a linear function of the continuum optical depth. The authors explained that the reason behind this choice is that they only observed a single line and they wanted to reduce the number of free parameters as much as possible. In our case, we do not have this limitation as we are going to analyse two spectral lines with different heights of formation observed with high spectral sampling. Thus, we can infer the atmospheric parameters in a wider range of heights with good accuracy. The second limitation in their work is the uncorrected contamination of the stray light. This stray light produces a lack of contrast in continuum images, and also affects to the size of the structures, and the amplitude of the magnetic Stokes profiles \citep{RuizCobo2013,QuinteroNoda2015,QuinteroNoda2016}. We solved the latter limitation applying a spatial deconvolution technique \citep{QuinteroNoda2015} that removes the stray light contamination induced by the spatial point spread function of the telescope. However, there is also a limitation from our side respect to the work of \cite{Sobotka2012}, and is the lack of high cadence observations because we use a slit spectrograph and, thus, we cannot track the evolution of the solar pore and the surrounding structures as they did in their work. Moreover, taking advantage of local correlation tracking techniques, these authors could infer the horizontal velocities in the surrounding area of the solar pore.

Finally, our aim, similar to \cite{Sobotka2012}, is to provide a comprehensive model of solar pores that help us to understand these structures and also provides a better input for numerical simulations. In addition, we also want to evaluate the possibilities and limitations of the spatial deconvolution method applied to this region because the Stokes profiles are affected by large horizontal intensity variations in small spatial scales, e.g., Stokes $I$ intensity goes from 0.5~$I_c$ (with $I_c$ the continuum signal) at the center of the pore to 1.1~$I_c$ at the neighbouring granulation in less than 3 arcsec, and this large variation could induce artefacts in the deconvolved data.

\section{Data analysis}

%\subsection{Observations}

We analysed in this work the magnetic properties of the active region NOAA 10949. It was detected for the first time by SOHO/MDI \citep{Scherrer1995} on 2007 March 28th as an individual dark spot in the continuum intensity images. The active region crossed the solar prime meridian on 2007 April 2nd, and lasted several days more, until 2007 April 4th. It was composed of a single solar pore during its lifetime and it showed low coronal activity. In addition, it was located relatively close to a filament although we cannot confirm if there is a connection between both structures due to the lack of data about the filament. We studied this solar pore using observations taken by the Spectropolarimeter ($SP$) \citep{Lites2013} on board $Hinode/SOT$ \citep{Kosugi2007,Tsuneta2008,Suematsu2008,Shimizu2008SOT} on 2007 April 1st between 18:56-19:57 UT. At this time, the active region was located at $(-150,220)$ arcsec, i.e. $\mu=0.96$. $Hinode/SP$ recorded the Stokes $I$, $Q$, $U$, and $V$ profiles for the Fe~{\sc i} 6301.5 and 6302.5 \AA \ spectral lines. It used a spectral sampling of 21.55 m\AA, and a pixel size and scanning step of about 0.16 arcsec providing a spatial resolution of 0.32 arcsec. The integration time was 4.8 s per slit position what produces a noise value of approximately $1\times10^{-3}$ of $I_c$ for the magnetic Stokes parameters \citep[for more information, see][]{Ichimoto2008,Lites2013SP_Prep}.

\begin{figure}
%\centering
\hspace{-0.2cm}
\includegraphics[width=9.0cm]{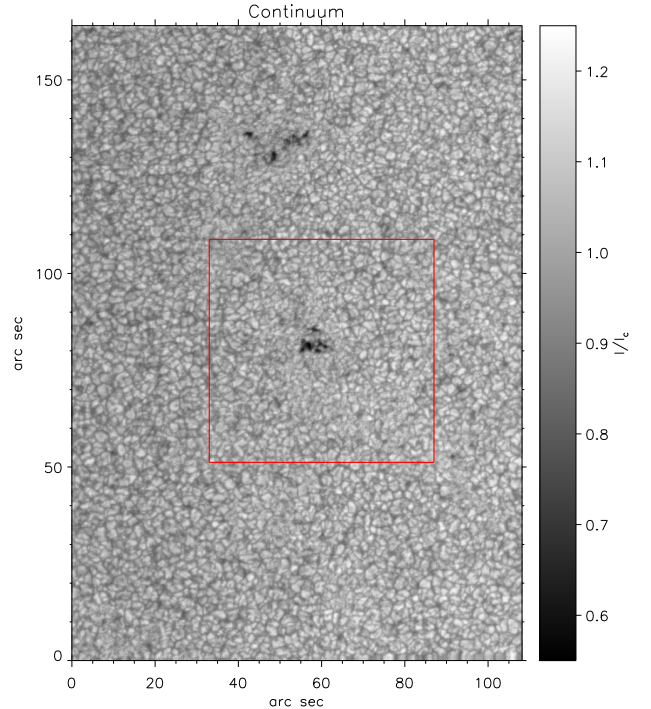}
\vspace{-0.4cm}
\caption{Continuum signal from the observation of 2007 April~1st 18:56 UT. Red square depicts the region we examine in detail in this work.}
\label{cont}
\end{figure}

\begin{figure*}
%\centering
\hspace{-0.3cm}
\includegraphics[width=18.0cm]{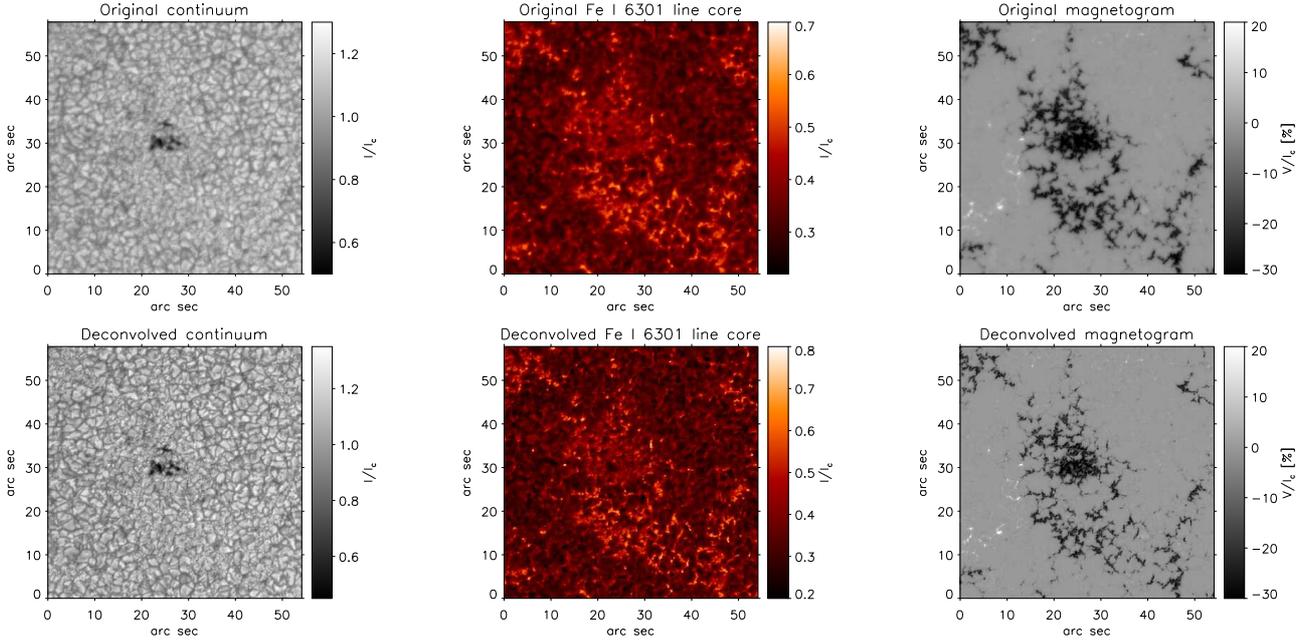}
\vspace{-0.5cm}
\caption{Comparison between original (upper row) and deconvolved (bottom row) observations. Different columns, from left to right, show the continuum intensity, line core intensity, and a Fe~{\sc i} 6302 \AA \ magnetogram.}
\label{compar}
\end{figure*} 

Figure \ref{cont} shows the continuum signal of the observed map examined in this work. We analysed a small fragment, marked in red, of the total observed map because we aim to focus only on the central pore. This active region is relatively small, with a size of around 10 arcsec, i.e. around 7 Mm, and it is composed of several small dots joined by brighter structures that resemble the granulation. It seems that the magnetic activity is not high enough to produce a large spot. In addition, after examining context Ca~{\sc ii} images taking by $Hinode/BFI$ we did not find any activity at upper heights either, being quiet during more than the three hours of observation examined.

\subsection{Deconvolution method}

We have performed the spatial deconvolution of the observed data using the same method presented in \cite{QuinteroNoda2015}. A detailed description of the procedure used in this work can be found in the mentioned paper. The number of principal components of eigenvectors we used to reconstruct the Stokes profiles is (8, 5, 4, 8) for ($I$, $Q$, $U$, $V$). This number of eigenvectors is similar to the one used on \cite{RuizCobo2013}, and \cite{QuinteroNoda2015} and smaller than the one used on \cite{QuinteroNoda2016}. This is because the examined region is close to disk center and the variety of Stokes profiles is smaller than the one found in the latter work. In addition, the predominately vertical nature of the magnetic field inside the solar pore demands just a few eigenvectors to reproduce the linear polarization profiles.

We performed 25 iteration steps in the deconvolution process. The original continuum contrast was 6.26 per cent and the value obtained after the deconvolution process is 9.73 per cent. We stopped the iteration process in this step because it provides an increase factor of the continuum contrast similar to the one obtained in previous works \citep{QuinteroNoda2015}.

\begin{figure*}
%\centering
\hspace{+0.2cm}
\includegraphics[width=16.0cm]{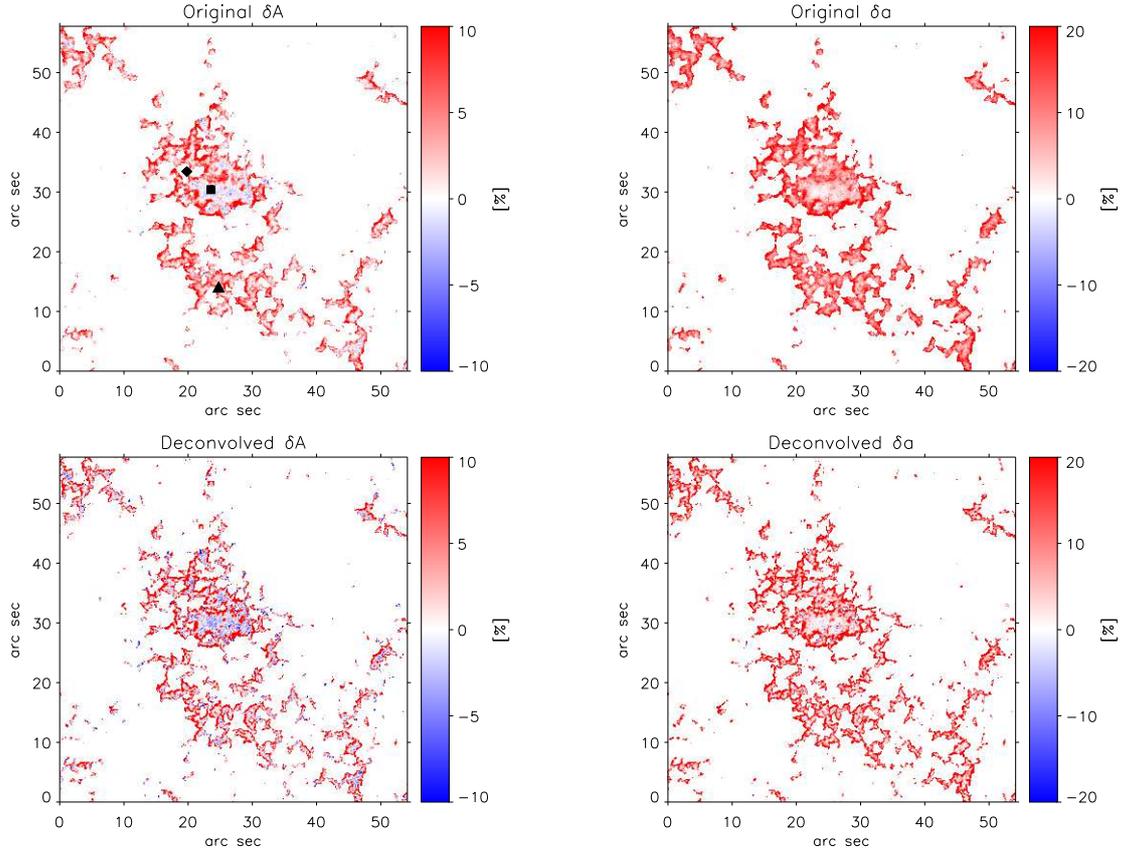}
\caption{Comparison between asymmetry maps from original (upper row) and deconvolved (bottom row) data. Left column displays the area asymmetry map while the amplitude asymmetry is depicted in the right column. White indicates null asymmetry while red and blue signifies positive and negative asymmetry, respectively. Symbols in top-left panel indicate the position of the profiles plotted on Figure \ref{per}.}
\label{as}
\end{figure*}

\subsection{Deconvolution results}

Figure \ref{compar} shows the comparison between the original (upper row) and deconvolved (bottom row) data. The field of view displayed in each panel corresponds to the region enclosed by the red square in Figure \ref{cont}.  We chose three different quantities to perform the comparison study: the continuum signal (leftmost column), line core intensity (middle column), and a magnetogram (rightmost column) built as the difference between Stokes $V$ maps taken at $\pm$100~m\AA \ from the line centre of Fe~{\sc i} 6302.5~\AA. As found in \cite{QuinteroNoda2015}, the continuum image looks sharper due to the increase in the continuum contrast. In addition, the abundance of intergranular bright points in the surroundings of the solar pore is more evident in the deconvolved map. 

If we examine the line core intensity, middle column, we are roughly measuring the temperature at upper layers, a few hundred kilometres above the continuum region. We can see that the pore is still dark at this height and surrounding it there is a web of bright structures that indicates higher temperatures than those of the surrounding quiet Sun granulation and correspond to the bright intergranular points seen in the continuum image. In addition, the magnetic structures located south of the pore forming a plage region also appear as hot structures at these layers. Comparing the original and deconvolved line core intensity panels we find more defined structures in the latter one and also with higher intensity values.

Finally, magnetogram panels (right column) show a single polarity structure that corresponds to the solar pore, and a surrounding plage with the same polarity. In this case, the deconvolved panel shows narrower structures with larger amplitude values. Remarkably, it is easier to see the separation between the magnetic substructures that formed the pore because the blurred effect present in the original data has disappeared.

\begin{figure*}
\centering
%\hspace{-0.5cm}
\includegraphics[width=17.3cm]{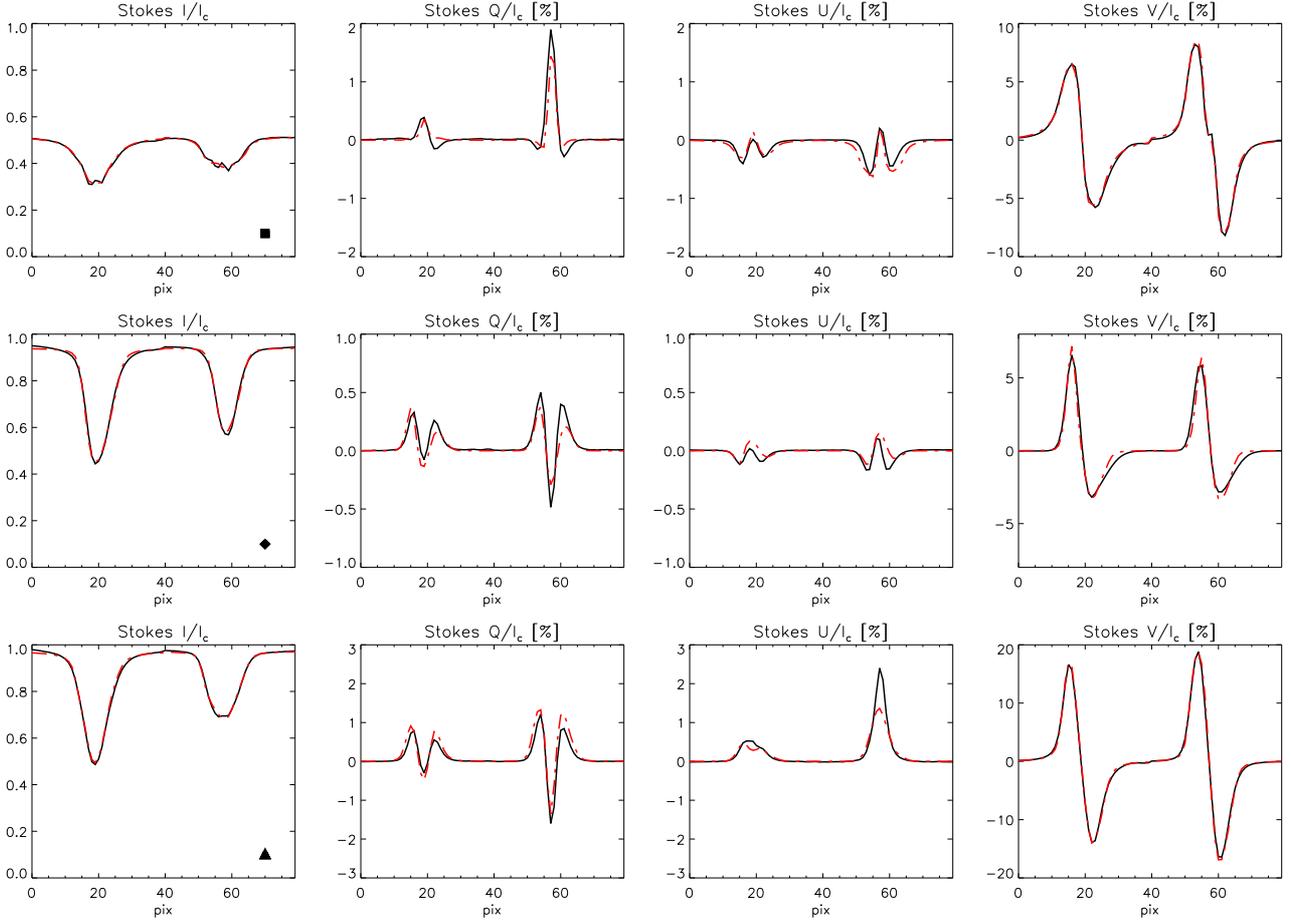}
\vspace{-0.3cm}
\caption{Stokes profiles from different regions in the observed map. Upper row shows a pixel from the central part of the pore, middle row from its edge, and bottom row displays a pixel from the solar plage. Black depicts the deconvolved profiles while red shows the results of the inversion. Symbols inside the panels of left column indicate their position in the observed map, see top-left panel of Figure~\ref{as}.}
\label{per}
\end{figure*}

\subsection{Analysis of Stokes profiles asymmetries}\label{asy}

As we mentioned before, a solar pore seems to be a simple structure composed of a vertical magnetostatic field. However, we want to analyze the amplitude and area asymmetries inside and at the edges of the pore to check if there is any type of pattern. In fact, we expect a change in the inclination of the magnetic field vector at these regions that could generate gradients along the line of sight, therefore asymmetries \citep{Illing1975}. We computed the asymmetries of the Stokes $V$ profile following the definition used in \cite{MartinezPillet1997}. The area asymmetry is obtained as

\begin{equation}
\delta A=s\frac{\sum_{i} V(\lambda_i)}{\sum_{i}\mid V(\lambda_i) \mid} ,
\end{equation}
where the sum is extended along the wavelength axis and $s$ is the sign of the Stokes $V$ blue lobe (chosen as $+1$ if the blue lobe is positive and $-1$ if the blue lobe is negative). The selected range of integration of the Stokes $V$ signals goes from $-0.43$ \AA\ to 0.43 \AA\ around the Fe~{\sc i} 6302.5 \AA\ line centre. Likewise, the amplitude asymmetry is defined as
\begin{equation}
\delta a=\frac{a_b-a_r}{a_b+a_r} ,
\end{equation}
where $a_b$ and $a_r$ are the unsigned maximum value of the blue and red lobe of Stokes $V$.

We computed these quantities over all the field of view enclosed by the red square in Figure \ref{cont} and we only considered pixels that show maximum Stokes $V$ amplitudes higher than $1\times10^{-2}$ of $I_{c}$. The results of this study are included in Figure \ref{as}. Top left panel shows that the area asymmetry is negative in the inner parts of the pore, dark in the continuum map, and with opposite polarity, positive, in the surrounding regions.  The reason behind these asymmetry changes between the central part of the structure and its edges could be a change in the line of sight component of the magnetic field being less inclined, i.e. almost vertical, in the core of the pore and more inclined in surrounding areas. Another explanation could be a change in the line of sight velocities that are usually small in the center of the structure and show large downflows at its edges \citep[for instance,][]{Morinaga2007,Sobotka2012}. We will delve into this aspect in following sections where we infer the atmospheric information performing inversions of the deconvolved Stokes profiles.

Contrary to the area asymmetries, amplitude asymmetries (right column) are always positive in the surroundings of the structure and close to zero in the inner part of the solar pore. In this case, some regions with negative amplitude asymmetry can be found in the deconvolved map although they are relatively scarce and with low amplitude. In that sense, we can confirm, as in \cite{QuinteroNoda2015}, that the deconvolution process barely modifies the Stokes $V$ profiles asymmetries.

\section{Inversion of Stokes profiles}

We obtain the physical information of the atmospheric parameters inverting the deconvolved Stokes profiles from the observed region enclosed by the red square of Figure \ref{cont}. We carried out the inversion of the Stokes profiles using {\sc sir} \citep[Stokes Inversion based on Response functions;][]{RuizCobo1992} code, which allows us to infer the optical depth dependence of the atmospheric parameters at each pixel independently.

\subsection{Configuration}

We are going to briefly summarize the configuration we used in this paper because is similar to the one used by \cite{QuinteroNoda2015,QuinteroNoda2016}. As the deconvolution process eliminates the stray light contribution of the spatial PSF, we do not need to include a stray light component in the inversion process. Consequently, we only used a single component for reproducing the observed profiles. The selection of nodes for each quantity is done using an automatic algorithm, based on the number of zeros of the response function of the corresponding quantity %\citep[see][]{QuinteroNoda2016b}.
(see Quintero Noda et al. 2016b, \textit{under revision})

In our case, to avoid too complex solutions, we limited the maximum number of nodes. We allowed five for temperature T($\tau$) (where $\tau$ refers to the optical depth evaluated at 5000 \AA), three for the line of sight (LOS) component of the velocity v$_{los}$($\tau$), three for the magnetic intensity B($\tau$), two for the inclination of the magnetic field $\gamma$($\tau$), one for the azimuthal angle of the magnetic field $\phi$($\tau$), and one for the microturbulence. The algorithm determines the optimum number of nodes for each atmospheric parameter at each iterative step, where this optimum number of nodes is always lower or equal to the maximum number of nodes.

On the other hand, macroturbulence is null and not inverted. At each iteration, the synthetic profiles are convolved with the spectral transmission profile of $Hinode/SP$ \citep{Lites2013}. Additionally, given that the inferred physical parameters could be reliant on the initial atmosphere, we minimize this effect by inverting each individual pixel with 10 different initial atmospheric models. These initial random atmospheric models were created as in \cite{QuinteroNoda2015}. Therefore, since each node corresponds to a free parameter during the inversion, our model could include up to 15 free parameters in case the automatic algorithm selects the maximum number of nodes allowed for the atmospheric parameters. 

Finally, the inversion code provides the uncertainties of the retrieved atmospheric parameters based on the response function to a given physical quantity. In that sense, if the response function is small, i.e. the Stokes profiles are not sensitive to changes in a given atmospheric parameter, the uncertainty of the solution for this parameter will be large. In addition, although the response functions depend on the atmospheric model itself \citep{Landi1977}, the height of formation of these lines only cover a small range of heights that usually goes from $\log \tau=0$ to $-2.5$. Therefore, we should restrict the analysis of the atmospheric parameters inside these general limits.

\subsection{Selected Stokes profiles}

We aim to examine the accuracy of the inversion at some spatial locations where the magnetic field configuration is complex. We chose a pixel that belongs to the inner part of the pore, a second one located at its edge, and a third pixel that belongs to the plage region that appears at the bottom part of the field of view (see Figure \ref{compar}). Deconvolved and fitted profiles are shown in Figure \ref{per}. If we start with the central part of the pore, upper row, we find that the fits are very accurate for Stokes $I$ and $V$, and relatively good for Stokes $Q$ and $U$. This is a tendency that is present for the rest of selected regions. We believe that this tendency produces fairly good results because the Stokes $V$ profiles, well fitted for almost all cases, display signals between 5-10 times larger than Stokes $Q$ and $U$ signals. Additionally, it is worth to mention the quality of the fitting for all Stokes $V$ profiles even when they show so dissimilar shapes, being with large widths and low amplitudes in the pore (upper row), with large area and amplitude asymmetries at its edges (middle row), or with extremely large amplitudes and moderate spectral redshifts as in the case of the solar plage (bottom row). Finally, we found a relatively good fit for the pixel belonging to the edge of the pore, middle panel, that could be related to the complex atmospheric stratification at the transition region between the pore and the surrounding granulation. To improve the accuracy of the fitting we would need more free parameters on the LOS velocity and magnetic field stratifications.

\subsection{Retrieved asymmetries}

\begin{figure}
%\centering
\hspace{+0.2cm}
\includegraphics[width=16.0cm]{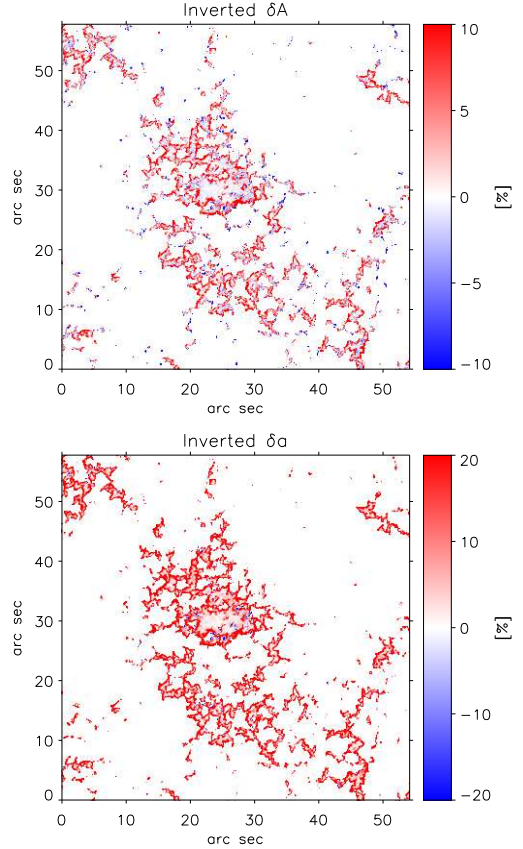}
\caption{Stokes $V$ area (top) and amplitude (bottom) asymmetries. They were computed using the fitted profiles from the inversion of the deconvolved data. The original asymmetries from the deconvolved profiles are shown on row of Figure \ref{as}.}
\label{as_inver}
\end{figure}

The Stokes $V$ profile asymmetries depend on velocity and magnetic field gradients. Therefore, an accurate match of these asymmetries serves as a criterion to 
check the reliability of the retrieved stratification of line of sight velocity and magnetic field. We plotted in Figure \ref{as_inver} the results for amplitude and area asymmetries obtained through the inverted profiles. If we compare these results with bottom panels of Figure \ref{as}, we can see that they are very similar, indicating that the accuracy of the fitting is high. However, there are minor differences in the central part of the pore, where the asymmetries are relatively smaller in the inverted profiles. We detected some pixels that can show a diminishing in area asymmetry from 13 per cent in the deconvolved profiles to 9.5 per cent in the inverted data. This reduction is smaller in the amplitude asymmetry although still present. However, we believe that these differences are sufficiently small to be certain that both, line of sight velocity and magnetic field gradients, are reliable.

\begin{figure*}
%\centering
\hspace{-0.5cm}
\includegraphics[width=18.3cm]{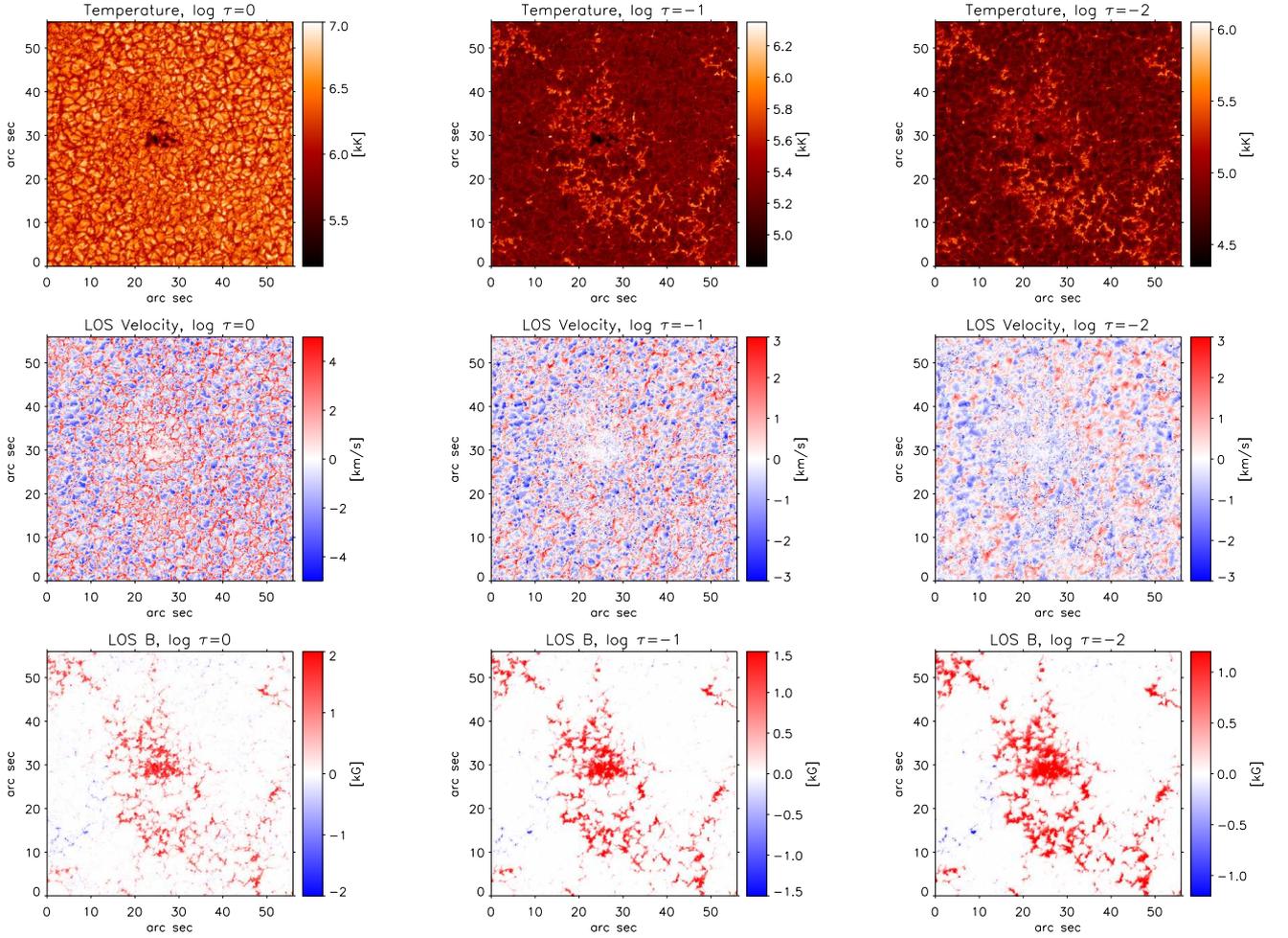}
\vspace{-0.3cm}
\caption{Spatial distribution of atmospheric parameters at different heights. From top to bottom row, temperature, LOS Velocity and, LOS magnetic field. From left to right column, optical depths $\log \tau=[0,-1,-2]$. Red LOS velocity indicates downflows while blue depicts upflows. White colors indicate a null velocity or a null longitudinal magnetic field.}
\label{inver_context}
\end{figure*}

\begin{figure}
\centering
%\hspace{-0.2cm}
\includegraphics[width=9.0cm]{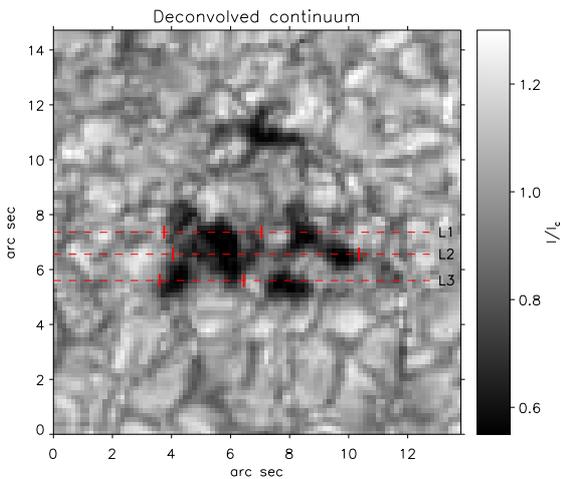}
\vspace{-0.4cm}
\caption{Continuum signal centred in the pore region. Horizontal lines indicate the location of the vertical cuts presented in Figure \ref{cuts}. Vertical ticks tentatively delimit the end of the pore substructures using the continuum map as reference.}
\label{context_small}
\end{figure}

\subsection{Spatial properties}

Figure \ref{inver_context} shows the spatial distribution of temperature, LOS velocity, and LOS magnetic field at different heights. Focusing first on the temperature, first row, we can see that its spatial distribution at $\log \tau=0$, leftmost panel, is similar to the continuum map, see Figure \ref{compar}, with hot granules, cool narrow regions, i.e. intergranules, and larger cool regions that conform the solar pore. If we move to upper layers, at $\log \tau=-1$ (middle panel), the landscape changes and resembles the line core information showed in middle column of Figure \ref{compar}. Plage regions are hotter than the surrounding quiet Sun while the pore is cooler. However, it seems that the height that is closer to the information displayed by the line core is $\log \tau=-2$, rightmost panel, because we can only see at this height a cool region at the left part of the pore structure, while the rest of the region shows temperatures similar to the quiet Sun, like the pattern found at line core wavelengths.

\begin{figure*}
\centering
%\hspace{-0.5cm}
\includegraphics[width=17.8cm]{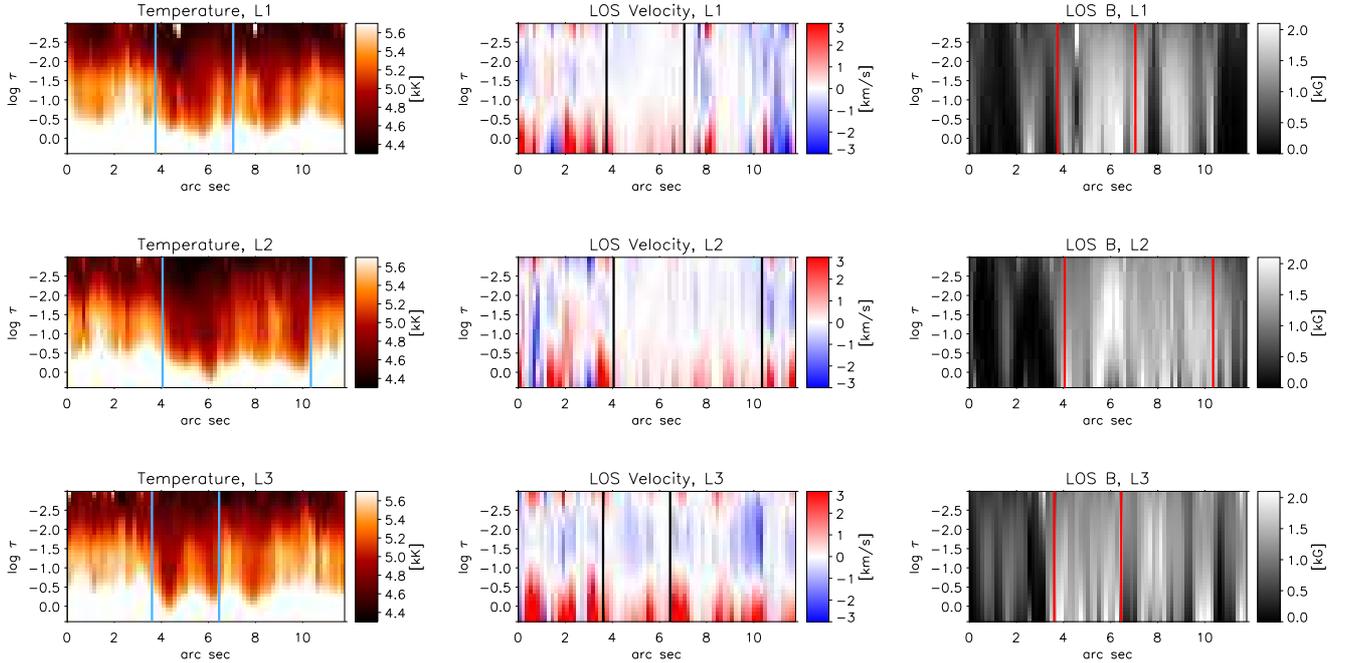}
\vspace{-0.3cm}
\caption{Vertical cuts of different atmospheric parameters corresponding to the horizontal lines displayed in Figure \ref{context_small}. Each row contains, from left to right, temperature, LOS velocity, and LOS magnetic field. We chose a different colour table than in Figure \ref{inver_context} for the latter physical quantity because, in this case, there is no change of sign of the LOS magnetic field in the examined regions being always with positive polarity or almost null. Vertical lines correspond to the location of the tick marks on Figure \ref{context_small}.}
\label{cuts}
\end{figure*}

Concerning the LOS velocity, the granulation pattern is seen at $\log \tau=0$ in quiet Sun regions while the inner part of the pore shows almost null velocities, white areas. In addition, there is a presence of strong downflows, red colour, at the periphery of the pore structure and at plage regions. If we analyse upper layers, middle and right panels, we find that LOS velocities slightly resemble the granulation pattern in quiet Sun areas although with lower velocities. In addition, the size of the structures is larger because they have expanded with height. However, the presence of the previously mentioned downflows detected at plage regions and at the edges of the pore have disappeared, indicating that these downward motions take place only at the bottom of the photosphere. We computed the height where these downflows velocities become null and we obtained a mean optical depth value of $\log \tau=-1.1\pm0.4$.

Finally, regarding the longitudinal magnetic field, we detected field strengths larger than 2~kG at the bottom of the photosphere, left panel, that decrease with height being in the range of $1-1.5$~kG at the middle photosphere, middle and right panels. Moreover, in this case, we can also detect an expansion of the magnetic structures with height.

\begin{figure*}
\centering
%\hspace{-0.5cm}
\includegraphics[width=16.0cm]{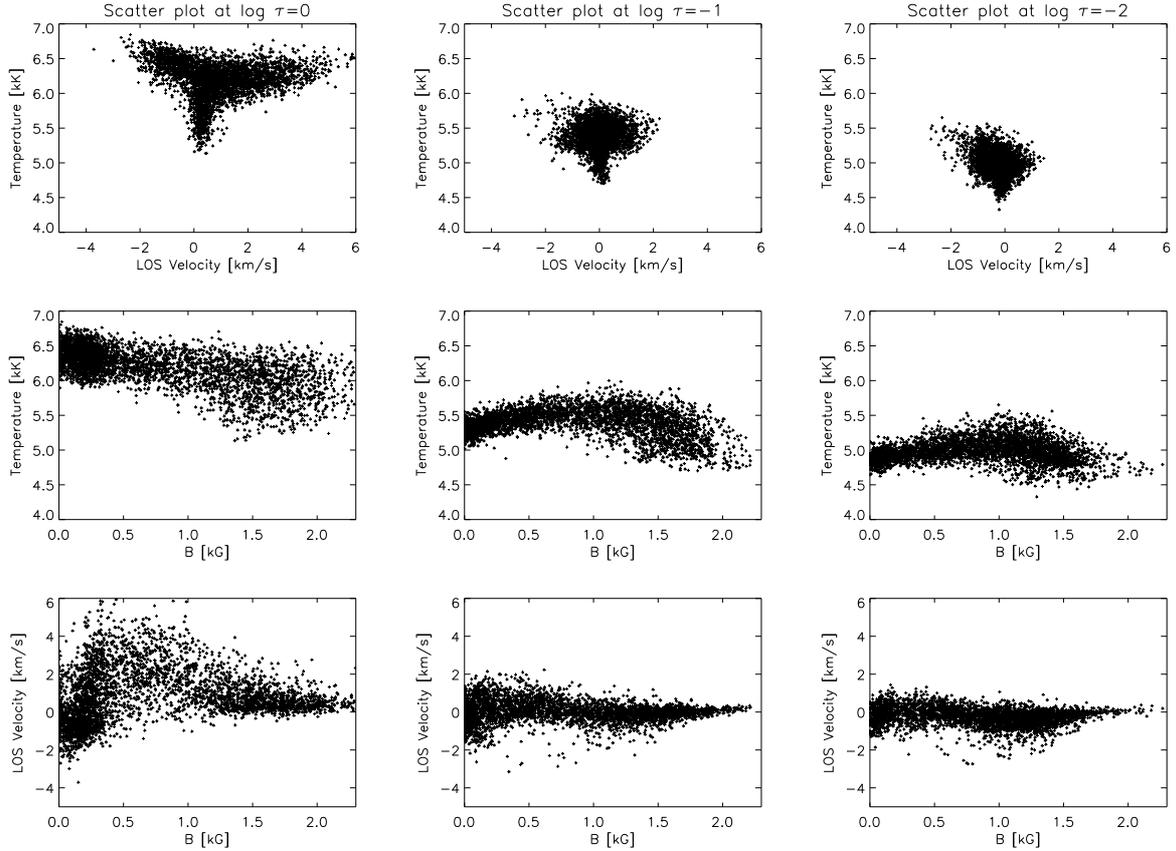}
\vspace{-0.3cm}
\caption{Scatter plots of three different atmospheric parameters; temperature, LOS velocity, and magnetic field strength. We show the spatial relation between these quantities at three different heights, i.e. $\log \tau=[0,-1,-2]$ (different columns). Different rows, from top to bottom, display the relation between temperature and LOS velocity, temperature and magnetic field strength, and LOS velocity and magnetic field strength.}
\label{scatter}
\end{figure*}

\subsection{Vertical cuts}

In order to study the optical depth stratification of the different physical quantities we selected three regions, see lines in Figure \ref{context_small}, that cross the pore. They are located at different positions in the south-north direction and have the same size in the east-west direction. We selected their location trying to examine the spatial differences between the substructures that compose the pore. The vertical stratification of the atmospheric parameters in these three regions is shown in Figure \ref{cuts} (see different rows). In addition, we added vertical lines in every panel that correspond to the vertical ticks of Figure \ref{context_small} in order to facilitate the visualization of pore substructures. We selected the location of these lines, or ticks in Figure \ref{context_small}, trying to delimited the end of the pore substructures using the continuum intensity as reference.

Starting with the temperature, first column, we can see strong fluctuations of its value at the base of photosphere, i.e., around $\log \tau=-1$, being at lower optical depths where the pore substructures are located (see the pixels in between the vertical lines). This property is present in all the different selected horizontal cuts (see different rows). The vertical stratification is relatively different between the inner part of the pore and the surrounding granulation, being hotter at middle heights (around $\log \tau=-1$) for the pixels outside the pore and cooler for the pixels belonging to the magnetic structure.

In the case of the LOS velocity, we also found a clear distinction between the pixels that belong to the pore structure and surrounding pixels. In the first case, we found null (or very small) LOS velocity values while, in the second case, we detected redshifted velocities around $3\sim4$ km/s at lower heights that go to zero or to small ($\sim1$~km/s) upflow velocities at middle layers (see the pixels near the vertical lines).

Regarding the line of sight magnetic field we found that is always positive in the selected areas, indicating that it does not change of polarity at the edges of the pore nor outside the magnetic structure. The magnetic field strength is large inside the structure with strengths around 2~kG that decreases with height up to $1$~$\sim$~$1.5$~kG at middle heights. Outside the magnetic structure, see vertical lines, the LOS magnetic field becomes almost null generating a large spatial contrast similar to the one observed in the continuum map (see Figure~\ref{context_small}).

Finally, we detected a large horizontal variation, even inside the pore substructures, in the three examined quantities. The reason could be that the pore substructures are constituted by a complex bundle of magnetic tubes that we are partially solving with the present spatial resolution, i.e. around 200 km. If we take a closer look of Figure~\ref{context_small}, we can see that the continuum intensity strongly fluctuates in the neighbouring granulation, but also in the pore substructures, being much darker at their central part than at their edges, located at just few pixels away.

\subsection{Spatial relations between atmospheric parameters}

We take advantage of the inversion results to examine the spatial relation between different atmospheric parameters, i.e. temperature, LOS velocity, and magnetic field strength, at different heights. In that sense, we aim to understand which is the relation between temperature and LOS velocity, temperature and magnetic field strength, and LOS velocity and magnetic field strength. We studied these relations in the pixels that belong to the solar pore as well as for its surrounding pixels. We selected a small region of $12\times12$ arcsec centred in the solar pore and displayed in Figure \ref{scatter} scatter plots between the mentioned atmospheric parameters at three different heights, $\log \tau=[0,-1,-2]$ (different columns in the figure).

We can start with the relation between the temperature and the LOS velocity (first row). At the bottom of the photosphere, left panel, we detect hot blueshifted pixels (negative velocities) that correspond to granules and slightly cooler redshifted pixels (positive velocities) that belong to intergranular lanes and to the pixels at the edges of the pore. We also see a set of pixels that covers a large range of temperatures and shows very low, or null, LOS velocities. We believe that the reason of this wide range of temperatures is that we have relatively hot pixels that can correspond to the interface between granules and intergranules and also very cool pixels that belong to the inner part of the solar pore. This large variation of temperature for similar velocities is the reason why the distribution of points in the scatter plot resembles the shape of the Greek character \textit{Y}. The same shape appears at higher heights but it is less clear as we move to upper layers because the pixels located at the edges of the pore now show almost null velocities. Lastly, we can see at all heights that blueshifted pixels usually correspond to hotter regions.

Middle row displays the relation between temperature and magnetic field strength. In this case, we can see at middle heights, where we are more sensitive to the magnetic field strength, a faint arcade shape indicating that the larger the magnetic field strength the cooler the temperature of the pixel (see also \cite{Sobotka2012}). In addition, we can also detect at middle layers, i.e. $\log \tau=-1$, that the pixels showing a large field strength (more than 1.5~kG) can reach temperature values as low as 4500~K.

Finally, bottom row displays the relation between LOS velocity and magnetic field strength.  Left panel, i.e. $\log \tau=0$,  reveals large positive velocities for intermediate magnetic field strengths that correspond to the pixels located at the edges of the pore. We can also detect very low (or null) velocities for the pixels with largest magnetic field strengths. These pixels are associated with the inner part of the magnetic structure. If we examine higher layers (middle and right panels) we see no trace of large downflows while the pixels harbouring a strong magnetic field strength still show almost null LOS velocities.

\section{Discussion and Conclusions}

We applied for the first time the spatial deconvolution technique on Hinode/SP observations of a solar pore. This region represents a challenge to the deconvolution code because it displays large continuum contrasts between the central part of the structure, with low intensities, and its surroundings, located at less than a few arcsec, that display the brighter granulation pattern. Thus, we faced this work with two main aims, the first was to test the capabilities of the spatial deconvolution code and the second was to analyse the physical properties of the magnetic structure. The first question reveals that the spatial deconvolution properly works without introducing ringing effects in the pore or artefacts in the Stokes profiles. In addition, the Stokes $V$ asymmetries are slightly modified by the process but only in their amplitude because the asymmetry sign does not change in most of the observed pixels. Moreover, we also found an improvement in the continuum contrast and a diminishing of the size of the magnetic structures in agreement with previous works on this topic.

Regarding the second aim of our study, i.e. the analysis of the physical properties of a solar pore, we performed inversions of the Stokes profiles in a large fragment of the original observed field of view. First, we checked that the retrieved atmospheric models lead to synthetic Stokes profiles that accurately fit the observed ones. These synthetic profiles even match the  Stokes $V$ amplitude and area asymmetries (although, in some cases, asymmetries were slightly lower in the inverted profiles), supporting the reliability of the retrieved line of sight gradients of the atmospheric parameters.

Later, we proceeded to examine the spatial distribution of the atmospheric parameters at different heights. We found that the inner part of the pore is cooler than its quiet Sun surroundings at all heights. At the same time, the neighbouring plage is hotter than the quiet Sun at middle layers. Then we examined the LOS velocity finding that is null in the inner part of the magnetic structure and that shows downward motions at its edges. These downward motions are only detected at lower heights, below $\log \tau\sim-1$, indicating that they are low photospheric features. Regarding the LOS magnetic field, we found strong field strengths, larger than 2~kG in the central part of the pore, and with single polarity everywhere. In addition, the magnetic field strength diminishes with height as the spatial size of the magnetic structures increases as we examine upper layers.

We also studied the vertical stratification of several cuts that cross the magnetic substructures of the pore. We found a significant decrease of the temperature inside the pore probably due to the inhibition of the overturning convection produced by its large magnetic field strength. In addition, the temperature between magnetic substructures was larger than inside the magnetic ones at middle heights. This contrast between magnetic and non magnetic substructures was also found in the LOS velocity. We detected null velocities at all heights inside magnetic concentrations and downward velocities outside them that can go to zero or turn to small upflow velocities at middle layers. Regarding the magnetic field strength, it shows maximum values around 2~kG at the low photosphere that decrease with height up to $1$~$\sim$~$1.5$~kG at middle layers. The inferred LOS magnetic field is unipolar in and outside magnetic substructures although its amplitude is very low outside magnetic areas generating a high spatial contrast between both regions. Finally, we examined spatial relations between temperature, LOS velocity, and magnetic field strength at different heights, corroborating all the properties described before for the pixels of the inner part of the pore as well as for the surrounding ones. 

We can conclude that the spatial deconvolution improves the overall quality of the observations without introducing any kind of artefacts and, at the same time, it allows us to infer the atmospheric information without relying on any type of stray light correction. In addition, our findings are in agreement with previous works as the ones presented by \cite{Morinaga2007}, \cite{Scharmer2011}, or \cite{Sobotka2012}. Moreover we also increase our knowledge about pores and their surroundings describing the height stratification of their atmospheric parameters. However, the examined structure does not show any kind of bright features, as bright dots or light bridges, that would be interesting to study after applying the spatial deconvolution method. In that sense, we need to check different solar pore regions to increase our knowledge about them. Additionally, this particular pore limits our study to the photosphere because it is quiet at chromospheric layers so we could not examine which type of photospheric activity can trigger an active chromospheric response.  Maybe it is due to its simple configuration but we plan to examine other solar pores trying to understand if the lack of chromospheric activity is a general property of solar pores or if it is just a particular condition of the present case.

\section*{Acknowledgements}
We thank A. Asensio Ramos his suggestions and ideas that helped to develop this work. BRC acknowledges financial support through the Project No. ESP2014-56169-C6-2-R funded by the Spanish Ministry of Economy and Competitiveness. We also thank the anonymous referee for providing helpful comments and suggestions that allowed to improve this work. \textit{Hinode} is a Japanese mission developed and launched by ISAS/JAXA, collaborating with NAOJ as a domestic partner, NASA and STFC (UK) as international partners. Scientific operation of the Hinode mission is conducted by the Hinode science team organized at ISAS/JAXA. This team mainly consists of scientists from institutes in the partner countries. Support for the post-launch operation is provided by JAXA and NAOJ (Japan), STFC (U.K.), NASA, ESA, and NSC (Norway).

\bibliographystyle{mnras} % style apj.bst
\bibliography{pore} % your references Yourfile.bib

% Don't change these lines
\bsp	% typesetting comment
\label{lastpage}
\end{document}